\documentclass[prl,twocolumn,showpacs,superscriptaddress,amsfonts,amsmath,
floatfix]{revtex4}
\usepackage{graphicx}
\usepackage{psfrag}
\usepackage{epsfig}
\usepackage{pstricks}
\usepackage{epstopdf} 
\newcommand{\Journal}[4]{#1 {\bf #2}, #3 (#4)}
\newcommand{\PRL}{Phys. Rev. Lett.}
\newcommand{\PRA}{Phys. Rev. A}

\newcommand{\Nature}{Nature}

\newcommand{\lt}{\left(}
\newcommand{\rt}{\right)}
\newcommand{\lqu}{\left[}
\newcommand{\rqu}{\right]}
\newcommand{\lgr}{\left\{}
\newcommand{\rgr}{\right\}}

\newcommand{\be}{\begin{equation}}
\newcommand{\ee}{\end{equation}}
\newcommand{\ba}{\begin{eqnarray}}
\newcommand{\ea}{\end{eqnarray}}
\newcommand{\fr}{\frac}
\newcommand{\nn}{\nonumber}
\begin{document}
\title {Detection of coherent superpositions of phase states by full counting statistics in a Bose Josephson junction}
\author{G. Ferrini}
\email{giulia.ferrini@grenoble.cnrs.fr}
\author{A. Minguzzi}
\email{anna.minguzzi@grenoble.cnrs.fr}
\author{F.W.J. Hekking}
\email{frank.hekking@grenoble.cnrs.fr}
\affiliation{Universit\'e Joseph Fourier, Laboratoire de Physique et Mod\'elisation des
Mileux Condens\'es, C.N.R.S. B.P. 166, 38042 Grenoble, France}
\date{\today}

\begin{abstract}
We study a Bose Josephson junction realized with a double-well potential. We propose a
strategy to observe the coherent superpositions of phase states occurring during the time
evolution after a sudden rise of the barrier separating the two wells. We show that their
phase content can be obtained by the full counting statistics of the spin-boson operators
characterizing the junction, which could be mapped out by repeated measurements of the
population imbalance after rotation of the state. This measurement can distinguish
between coherent superpositions and incoherent mixtures, and can be used for a
two-dimensional, tomographic reconstruction of the phase content of the state.
\end{abstract}
\maketitle

\section{I. Introduction}

Superpositions of quantum states are at the heart of quantum physics. They are important
for precision tests and are at the basis of quantum information and communication. The
fate of these superpositions in the presence of interactions with the environment has
been largely studied to understand the cross-over from a quantum to a classical description \cite{zureck}.
Although fragile and subject to decoherence, superposition states have been
experimentally observed with trapped ions \cite{leibfried05} and cavity photons
\cite{deglise}.

Ultracold atomic gases are another system potentially suitable for realizing macroscopic
superpositions. We are interested here in the study of superpositions of
(angular-momentum) coherent states \cite{Arecchi}, which could be realized in a Bose
Josephson junction made by a double-well trap. Such superposition states are predicted to
form \cite{yurke_stoler} during time evolution after a sudden rise of the barrier
separating the two wells, starting from a coherent state at the initial time. An
initially coherent state corresponds to the ground state in the regime where the
tunneling energy dominates the repulsion energy, and is commonly realized in
current experiments on Bose Josephson junctions \cite{expbosejj}. The subsequent
dynamical evolution after the rise of the barrier is driven by interactions only, and  at
specific times the system undergoes the formation of several multicomponent
superpositions of phase states, finally coming back to the initial state after a revival
time. In a pioneering experiment \cite{Bloch2002b} on ultracold bosons in optical
lattices, revivals have been indeed observed, but it was not possible to  directly access
the macroscopic superpositions which were formed during intermediate stages of the
evolution.

Multicomponent superpositions of phase states are characterized by a sequence of equal
and equidistant peaks along the equator $\theta=\pi/2$ of the Husimi phase distribution
$Q(\theta, \phi)=\langle \theta \phi |\hat \rho |\theta \phi \rangle$ \cite{husimi,noi},
with $|\theta \phi\rangle$ an angular-momentum coherent state and $\hat \rho$ the density
matrix. Each peak corresponds to the contribution of a given coherent state belonging to
the coherent superposition. An important issue to address is how to reveal the effect of
decoherence on quantum superpositions of macroscopically distinguishable states. We
proceed by analogy with the macroscopic superpositions of phase states studied in the
context of photons in cavities \cite{Haroche_book}. We expect that generic sources of
noise would tend first to destroy the coherence between the components of the
superposition, by projecting it onto an incoherent mixture of coherent states. On a longer
time scale each component is expected to relax, yielding a total spreading of the phase
profile.

\begin{center}
\begin{figure}
%
\includegraphics*[width=0.9\columnwidth]{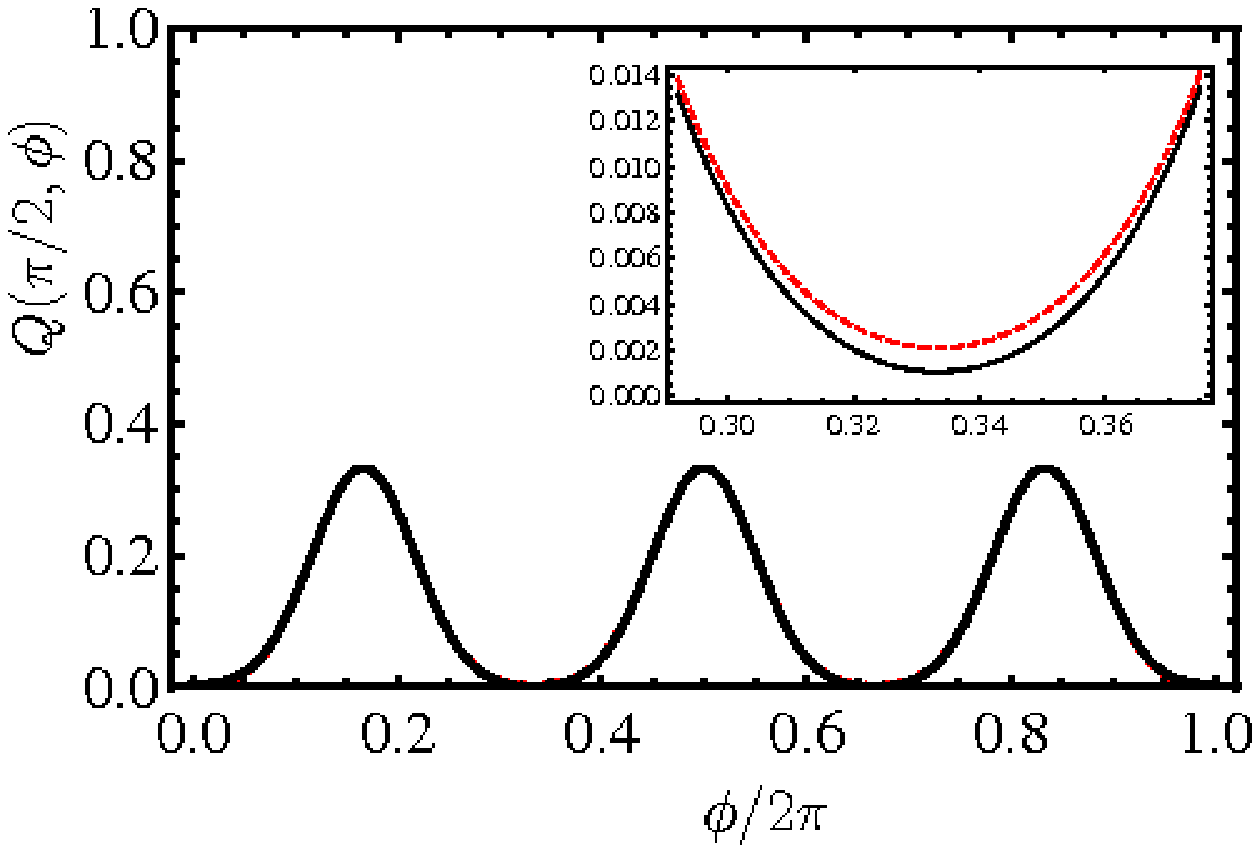}
\includegraphics*[width=0.7\columnwidth]{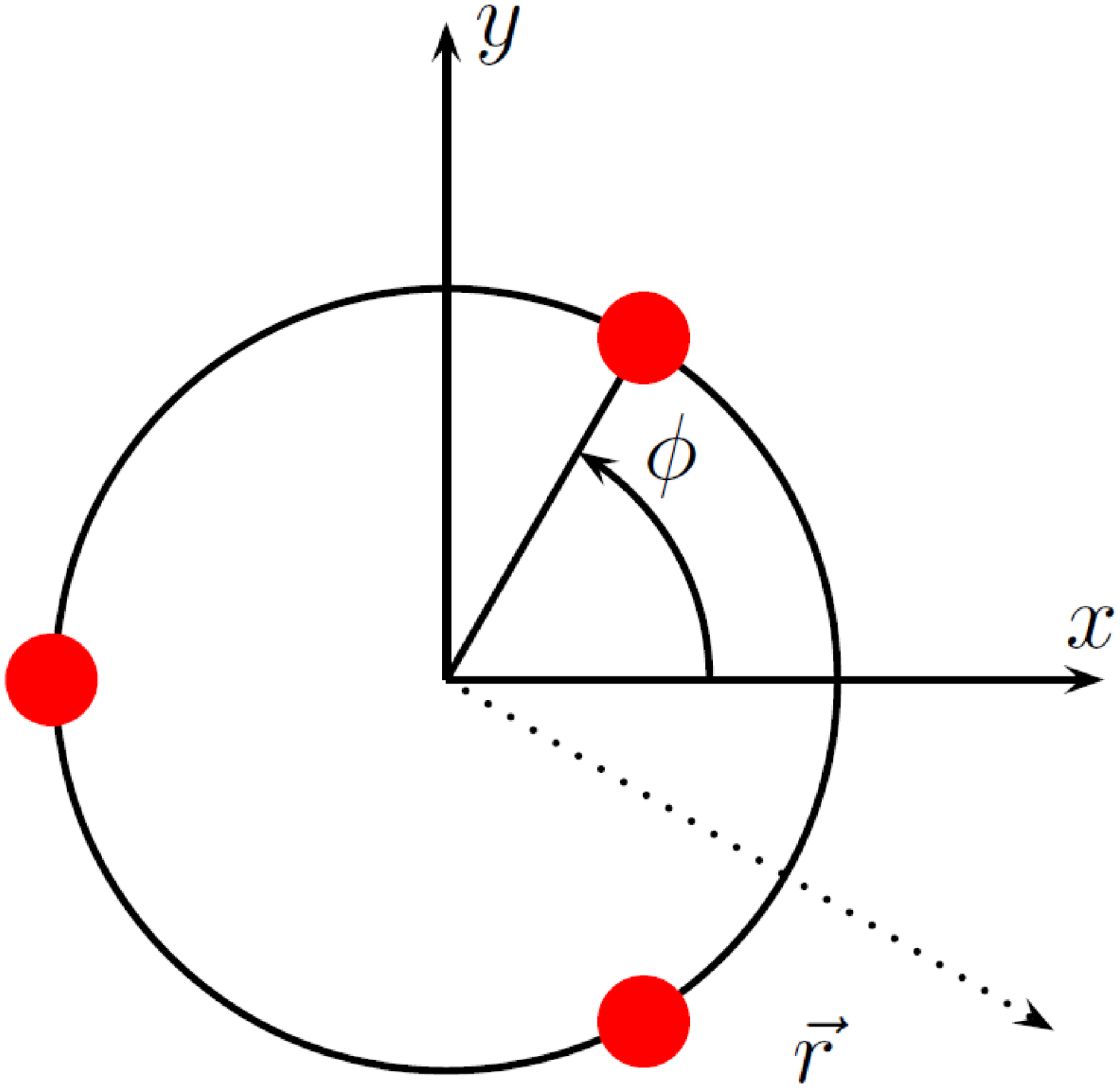}
%
\caption{(Color online) Top panel:  Dimensionless Husimi distribution $Q(\theta =
\pi/2,\phi)$ for a three-component superposition of phase states (solid line) and for the
corresponding incoherent mixture (dashed line), as a function of the phase $\phi/2\pi$
for $N$=20 particles. The inset shows a zoom of the same function around $\phi = 2 \pi/3
$, illustrating the difference between the superposition state and the incoherent
mixture. Bottom panel: section of the Bloch sphere on the equatorial plane $\theta =
\pi/2$, parametrized by the angle $\phi$. The dots indicate schematically the phase of
the three coherent states which give rise to the superposition, and correspond to the maxima in the Husimi distribution illustrated in the top panel. The vector $\vec r$
defines a generic direction of the spin-boson operator for which the probability
distribution is considered (see text). Our convention for the $x,y$ axes is also
indicated. } \label{fig1}
\end{figure}
\end{center}

In this work we address the question on how to identify the quantum macroscopic
superposition of phase states and how to distinguish them from mixtures. Our approach is
substantially different from the one of Ref.~\cite{Piazza08} which is devoted to map out
the Husimi distribution. Although the Husimi phase distribution is in one-to-one
correspondence with the full density matrix \cite{husimi}, in practice it is almost
insensitive to the difference between a coherent superposition of phase states and the
corresponding incoherent mixture. This is because the Husimi distribution is the diagonal
of the density matrix represented over coherent states. Figure \ref{fig1} (top panel)
shows the Husimi distribution for a three-component superposition of phase states and for
the incoherent mixture, the tiny difference between the two being illustrated in the
inset.

The paper is organized as follows. After introducing the model in Sec. II, in Sec. III we
propose to reconstruct the phase distribution by mapping out all the higher order
correlations, ie the full counting statistics of the spin-boson operators $\hat J_x$ and
$\hat J_y$ associated to the quantum model of the junction. In Sec. IV we show that it is
possible to make a two-dimensional reconstruction of a Wigner-like distribution function.
We conclude in Sec. V after discussing experimental issues.

\section{II. Two-mode model for the Bose Josephson junction}

We describe the Bose Josephson junction in the
quantum regime by using the two-mode approximation. This model applies in the limit where
the chemical potential of the system is smaller than the barrier between the two wells
and corresponds to neglecting higher order single-particle levels of the double-well
potential. Hence we adopt the following two-mode Bose-Hubbard Hamiltonian:
\begin{eqnarray}
\label{Ham2mode}
H=E_1  \hat a_1^\dagger \hat a_1 +E_2  \hat a_2^\dagger \hat a_2
+\frac{U_1}{2}\hat a_1^\dagger \hat a_1^\dagger \hat a_1 \hat a_1 \nonumber \\
+ \frac{U_2}{2}\hat a_2^\dagger \hat a_2^\dagger \hat a_2\hat  a_2
- K(\hat a_2^\dagger \hat a_1 + \hat a_1^\dagger \hat a_2),
\end{eqnarray}
where $\hat a_i,\hat a_i^\dagger$ with $i=1,2$ are bosonic field operators for the bosons
in each well, satisfying the usual commutation relations $[\hat a_i,\hat
a_j^\dagger]=\delta_{ij}$; $E_i $ are the energies of the two wells, $U_i>0$ are the
boson-boson repulsive interactions and $K$ is the tunnel matrix element, ie the Rabi
oscillation energy in the case of a non-interacting model. We then perform the
transformation onto spin-boson variables, where $\hat J_x=(\hat a^\dagger_1 \hat a_2+
\hat a^\dagger_2 \hat a_1)/2$ is the inter-well tunneling operator, $\hat J_y=-i (\hat
a^\dagger_1 \hat a_2-\hat a^\dagger_2 \hat a_1)/2$ is the current operator, and $\hat
J_z=(\hat a^\dagger_1 \hat a_1- \hat a^\dagger_2 \hat a_2)/2=\hat n$ is the population
imbalance operator.  The Hamiltonian is readily rewritten as
\begin{equation}
\hat H=U_s (\hat J_z-n_0)^2/2 -2 K \hat J_x,
\label{angham}
\end{equation}
with $U_s=(U_1+U_2)$ and $n_0$ is related to the imbalance between the wells according to
$n_0=-[E_1 +U_1(N-1)/2-E_2 -U_2(N-1)/2]/U_s$ \cite{noi}.
In the quasi-classical limit $KN\gg U_s$, the angular-momentum coherent states $|\theta_0
\phi_0\rangle=\sum_{m=-N/2}^{N/2} \left(  \begin{array} {c} N \\ m+N/2
\end{array}\right)^{1/2} \frac{\alpha^{m+N/2}}{(1+|\alpha|^2)^{N/2}} |m\rangle$ with
$\alpha=\tan(\theta_0/2) \exp(-i\phi_0)$ are useful to describe the state of the system. In particular, the ground state corresponds to
the coherent state with $\theta_0=\pi/2$ and $\phi_0=0$.

In the following we consider the time evolution of the Hamiltonian (\ref{angham}) after a
sudden rise of the barrier, namely we set $K=0$,  starting from the noninteracting ground
state $|\theta_0=\pi/2,\phi_0=0\rangle$ as the initial state. Such a dynamical evolution
occurs entirely in the equatorial plane of the Bloch sphere as $\hat{J}_z$ is a constant
of the motion. We focus on the specific time $T_q=2 \pi/(U_s q)$ where a coherent
superposition of $q$ coherent states is formed, according to \be |\psi_q\rangle  =
\tilde{u}_0 \sum^{q - 1}_{k = 0} \tilde{c}_k |e^{ - i\fr{2 \pi k}{q} - i\fr{\pi}{q}}
\rangle \ee with $\tilde{u}_0 = \fr{1}{q} e^{\fr{i \pi N}{2 q}} \sum_{m=0}^{q - 1} e^{-
i\fr{\pi m(m - 1) }{q}}$ and $\tilde{c}_k = e^{ i\fr{\pi k (N + k + 1)}{q}}$; we have
assumed $q$ odd. Our aim is to reveal the phase content of such a state. For this purpose
we consider the operators $\hat J_x$ and $ \hat J_y$ which have an action close to that
of the relative-phase operators $\hat{\cos\phi}$  and $\hat{\sin\phi}$. However, the
latter act incorrectly on the boundary states $|\pm N/2\rangle$. The average over a
coherent state characterized by a relative phase $\phi_0$ and population imbalance $n$
gives $\langle\alpha |\hat J_x|\alpha\rangle= n \cos\phi_0$ and $\langle\alpha |\hat
J_y|\alpha\rangle= n \sin\phi_0$. While the momentum distribution accesses only the
first-moment averages $\langle\hat{\cos\phi}\rangle$, $\langle\hat{\sin\phi}\rangle$
\cite{Stringari_Pitaevskii}, in order to access  the entire phase distribution we need to
consider all the higher correlations.

\section{III. Full counting statistics of the angular-momentum operators}

The probability
distribution of the eigenvalues of an observable contains the information equivalent to
the knowledge of all the moments of the distribution itself. In turn, the latter are
known once the generating function is known, according to a full counting statistic
approach \cite{Reichl}. 

Let us consider a spin-boson operator $\hat J_r$ in the direction
specified by the two-dimensional vector $\vec r=(\sin \phi, -\cos\phi)$ in the $(x,y)$
plane. The generating function for the probability distribution of the eigenvalues $r$
that correspond to the eigenstates $|r\rangle$, $\hat J_r |r\rangle = r |r\rangle$, is
given by
\begin{equation}
h_\phi(\eta)=\langle e^{-i\eta \hat J_r} \rangle. \label{eq:generating}
\end{equation}
Here, $\hat J_r=\hat J_x \sin \phi -\hat J_y \cos \phi$ and $\langle... \rangle$
indicates the quantum average over the state of the system. Considering for simplicity a
pure state $|\psi\rangle$ and expanding in terms of the eigenstates $|r\rangle$ of $\hat
J_r$ we have
\begin{equation}
\label{eq:genfun}
h_\phi(\eta)=\sum_r e^{-i \eta r } |\langle \psi| r\rangle|^2,
\end{equation}
where the eigenvalue $r$ takes integer values in the interval $\lqu -N/2, N/2 \rqu$ and
we have assumed $N$ to be even. We shall denote $P_\phi(r)= |\langle \psi| r\rangle|^2$
the corresponding probability distribution (for a generic density matrix $\hat{\rho}$,
$P^{\hat{\rho}}_{\phi}(r) = {\rm Tr}(\hat{\rho}|r \rangle \langle r |)$). Its shape
reflects the phase content of the state projected along the direction specified by the
vector $\vec r$ (see the bottom panel of Fig.\ref{fig1} for a sketch). As we will detail in Sec.V, this could be accessed experimentally.

In order to determine the probability distribution $P_\phi(r)$, we evaluate the
generating function Eq.(\ref{eq:generating}) analytically both for coherent
superpositions and mixtures. Specifically, we focus on the three-component superposition
of phase states and on the corresponding incoherent mixture of the same three equally-weighted phase
states. We use the definition of angular-momentum coherent states and
the disentangling formula \cite{Arecchi}
\begin{equation}
\label{eq:disaccoppiamento} e^{-i \eta \hat{J}_r}= e^{- \tau^* \hat{J}_-}e^{- \log ( 1 +
|\tau|^2 ) \hat{J}_z} e^{ \tau \hat{J}_+},
\end{equation}
with $\tau=\tan(\eta/2) e^{-i\phi}$. We thus obtain the final results for the generating
functions: \ba
&& h^{mixt}_\phi(\eta) = |\tilde{u}_0|^2 \sum_{k = 0}^{q - 1}  \\
&& \lgr |\cos \fr{\eta}{2}| + i \sin \fr{\eta}{2} \, \, {\rm sign} \lqu \cos \fr{\eta}{2}
\rqu \sin\lt \fr{2\pi k}{q} + \fr{\pi}{q} - \phi \rt \rgr^N  \nn \ea for the incoherent
mixture and \ba
&& h^{coh}_\phi(\eta) =  h^{mixt}_\phi(\eta) \\
&& +|\tilde{u}_0|^2 \sum_{k \neq k' = 0}^{q - 1} \fr{\tilde{c}_k \tilde{c}_{k'}^*}{2^N}  \lgr |\cos \fr{\eta}{2}| \lt 1 + e^{-i\fr{2 \pi(k - k')}{q}} \rt \right. \nn \\
&& \left. + \sin \fr{\eta}{2} \, \, {\rm sign} \lqu \cos \fr{\eta}{2} \rqu \lt e^{i
(\fr{2 \pi k'}{q} + \fr{\pi}{q} - \phi)} - e^{-i (\fr{2 \pi k}{q} + \fr{\pi}{q} - \phi)}
\rt  \rgr^N \nn \ea for the coherent superposition. The probability distribution
$P_\phi(r)$ is readily obtained by Fourier transforming the above expressions.

\begin{center}
\begin{figure}
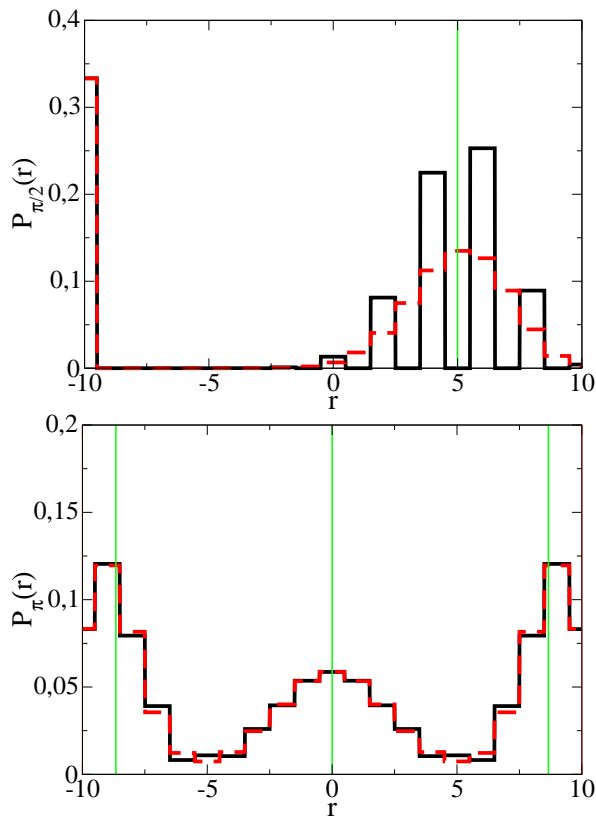

%
\psfrag{P(nx)}[]{$P_{\pi/2}(r)$}
\psfrag{nx/(M/2)}[]{$r$}
\includegraphics*[width=0.9\columnwidth]{fig2a.eps}
\psfrag{P(ny)}{$P_\pi(r)$}
\psfrag{ny/(M/2)}{$r$}
\includegraphics*[width=0.9\columnwidth]{fig2b.eps}
%
\caption{(Color online) Eigenvalue distribution $P_\phi(r)$ corresponding to $\hat{J}_x$
($\phi=\pi/2$) and $\hat{J}_y$ ($\phi=\pi$) for the three-component coherent
superposition (solid lines) as well as for the incoherent mixture of the same phase
states (dashed lines) with $N=$ 20. The vertical lines correspond to the semiclassical
values for $\langle \hat{J}_x\rangle$ and  $\langle \hat{J}_y\rangle$ for the coherent
states entering the superposition.} \label{fig2}
\end{figure}
\end{center}

By considering the projections along the $x$ and $y$ directions, respectively, we obtain
the full counting statistics of the operators $\hat J_x$  and $\hat J_y$. This is illustrated in
Fig.\ref{fig2} for the three-component superposition $|\psi_3\rangle$ of phase states as
well as for the corresponding mixture. The distribution is peaked around the
semiclassical values for $\langle\hat J_x\rangle=(N/2) \cos(\pm \pi/3)$, $(N/2)
\cos(\pi)$ and $\langle\hat J_y\rangle=(N/2)\sin(\pm \pi/3)$, $(N/2) \sin(\pi)$. We find
that the distribution $P_{\pi/2}(r)$ displays a noticeable difference between the mixture
and the coherent superposition: the latter displays oscillations which are absent in the
former. The presence of fringes in the distribution of the eigenvalues of angular
momentum operators for superposition states was also noticed in the context of the
dynamics of the quantum non-linear rotator by Sanders \cite{Sanders}. The function
$P_\pi(r)$ instead does not display fringes for the three-component superposition because its
components do not overlap when projected along the $y$-direction (see Fig.\ref{fig1},
bottom panel); as a result no interference effect takes place in this case.

The result for the three-component state extends to higher-component superpositions; the
two-component one instead cannot be distinguished from the corresponding incoherent
mixture by this method, due to the specific form of its state components. The full
 counting statistics of the operator $\hat J_z$ could also be defined, but does not yield
any useful information about the considered superpositions of phase states as it
coincides with the binomial distribution $P_{\phi = 0}(r) = \fr{1}{2^N}{N \choose
\fr{N}{2} + r}$ of the initial coherent state.

\section{IV. Two-dimensional reconstruction of the phase distribution}

A two-dimensional (2D) tomographic reconstruction of the phase content of a state in
the $(x,y)$ plane is possible, using the concept of the Radon transform
\cite{Radon_transform}. This has been also used for coherent superpositions with cavity
photons \cite{deglise}. The 2D distribution function $f(x,y)$ is obtained once all the
one-dimensional projections $P_\phi(r)$ are known as a function of $\phi$ in the
interval $[0,2\pi]$. One should invert the expression \be P_\phi(r)=\int_{- \infty}^{+ \infty} dx \int_{- \infty}^{+ \infty} dy
f(x,y) \delta(r-x \sin\phi + y \cos\phi), \ee which, using the definition of the
generating function in Eq.~(\ref{eq:genfun}), yields as a final result \be f(x,y)=
\frac{1}{(2\pi)^2}\int_0^\pi \eta d\eta \int_0^{2\pi} d\phi\, h_\phi(\eta) e^{i \eta (x
\sin\phi -y \cos\phi)}. \ee This is a quasi-probability distribution for the non-commuting
operators $J_x$ and $J_y$. It could be regarded as the two-dimensional projection on the
equatorial plane of the Bloch sphere of a SU(2) Wigner function in the spirit of
Ref.\cite{Chumakov}. It is readily verified that the marginal probability distribution
obtained by integration of $f(x,y)$ along one of the two directions $y$ or $x$ yields the
one-dimensional distributions $P_{\phi}(r)$ with $\phi=\pi/2$ and $\pi$, respectively.
The function $f(x,y)$ is not in one-to-one correspondence with the state of the system.
However, for the specific superpositions of phase states which we consider here it yields
the main information about the phase structure of the state. Furthermore, in the case of
a coherent superposition of phase states it is markedly different from the corresponding
incoherent mixture.

\begin{center}
\begin{figure}
\includegraphics*[width=\columnwidth]{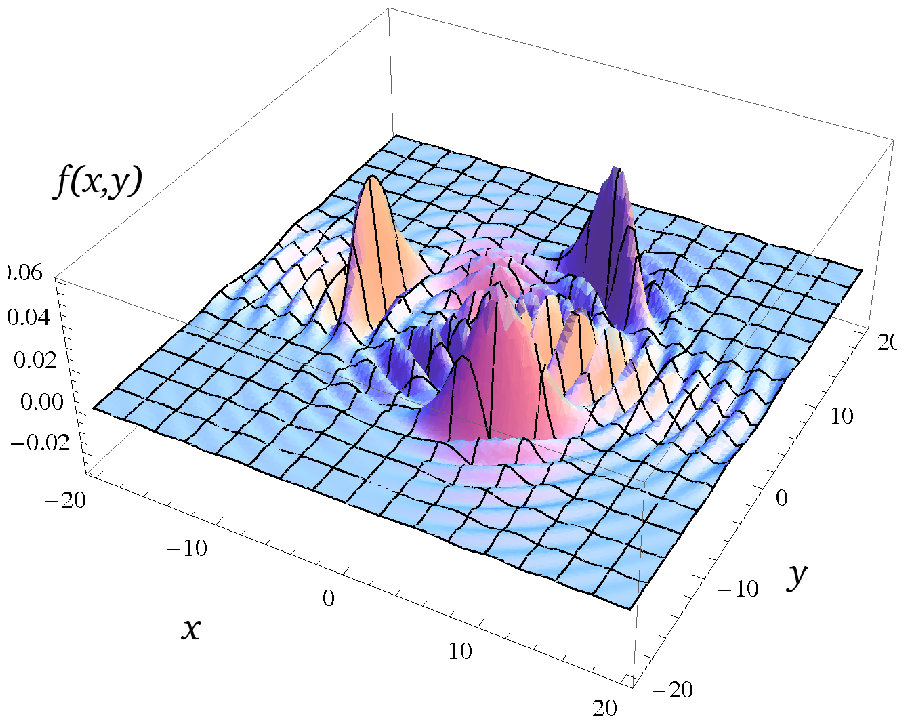}
\includegraphics*[width=\columnwidth]{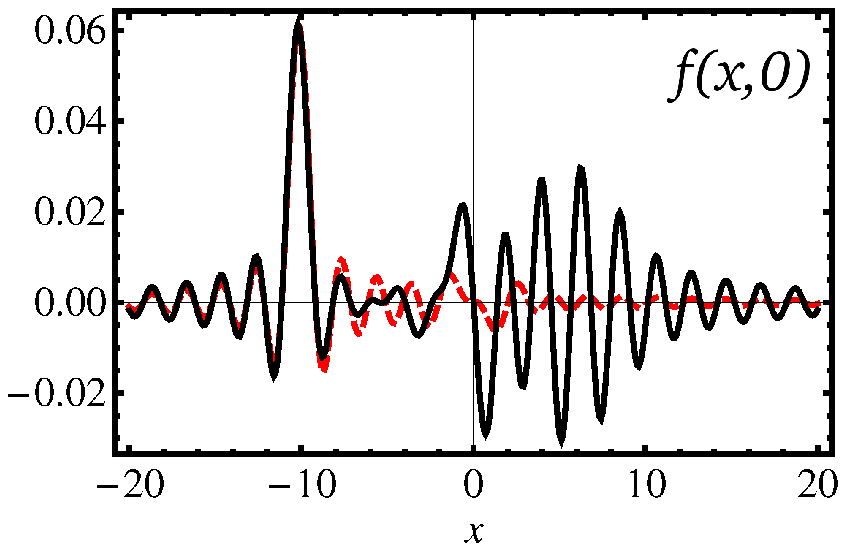}
\caption{(Color online) Top panel: Dimensionless 2D quasi-probability distribution $f(x,y)$ in
the $(x,y)$ plane (dimensionless) for a three-component coherent superposition with
$N=20$ particles. Bottom panel: section of the same quasi-probability distribution (solid line) in the same
units, in the direction $y=0$. The dashed line represents the quasi-probability function $f(x,0)$ 
for the corresponding incoherent mixture of the same three coherent states.}
\label{fig:2D}
\end{figure}
\end{center}

Figure~\ref{fig:2D} illustrates the 2D quasi-probability distribution $f(x,y)$ for a
three component coherent superposition. It shows three pronounced maxima in
correspondence with the three coherent states giving rise to the macroscopic
superposition. It also displays oscillations between the maxima, due to interferences
between the components. The 2D quasi-probability function evaluated for the corresponding
incoherent mixture also exhibits the main peaks but the fringes are strongly suppressed
(see Fig.\ref{fig:2D} bottom panel), the small remaining oscillations  being intrinsically
due to the definition of the function $f(x,y)$ as a Fourier transform in angular
variables.

\section{V. Experimental issues and concluding remarks}

For each choice of the angle $\phi$ the probability distribution $P_{\phi}(r)$ can be
experimentally accessed by repeated measurements of the corresponding angular momentum
operator $\hat{J_r}$. Indeed, since the eigenstates of $\hat{J_r}$ form an orthonormal
basis, each superposition state decomposes as $| \psi_{q} \rangle = \sum_{r = -
N/2}^{N/2} c^q_{r} | r \rangle$  with $ c^q_{r} = \langle r |\psi_{q} \rangle$. Then,
according to the postulates of quantum mechanics, after a (projective) measurement of
$\hat{J_r}$ the state jumps to the state $| r \rangle$ with probability
$P_{\phi}^{coh}(r) = |c^q_{r}|^2 $, and the corresponding outcome of the measurement is
$r$. The full distribution $P_{\phi}(r)$ is obtained by repeating this procedure many
times, each time preparing the system in the same initial state~\cite{nota}. The
measurement of $\hat{J}_r$ for a generic angle $\phi$ can be achieved by measuring the
population imbalance $\hat{J}_z$ between the two wells -- a variable typically accessed
in experiments  \cite{expbosejj} -- after proper rotations of the state over the Bloch
sphere \cite{Kim_1998}.  More precisely, measuring $\hat{J}_r$ on the quantum state
$|\psi \rangle$ is achieved by measuring $\hat{J}_z$ on the rotated state $e^{i \pi
J_x/2} e^{i\phi J_z} |\psi \rangle$. Each rotation can be implemented by a time evolution
of the system with Hamiltonian (\ref{angham}) once the interaction strength $U_s$ is
tuned to zero (eg by a Feshbach resonance). The rotation direction and angle can be
chosen by tuning the height of the barrier, the energy bias between the wells and the
time-evolution interval \cite{moore}. 

In summary, using the concepts of full counting
statistics we propose a method to access to the full phase content of a macroscopic
superposition of phase states, as well as to reconstruct tomographically its
two-dimensional quasi-distribution. This method is capable to distinguish between such
states and an incoherent mixture as the latter does not display fringes in the
probability distribution. The superposition of phase states could be realized with
current experiments on Bose Josephson junctions.

Feasibility of the current proposal requires control of the atom number on the
experiment; atom losses  are predicted to destroy the superposition and seem to present
the major experimental challenge.

During the preparation of the present manuscript we have became aware of similar results
for the 'NOON' state by Haigh et al. \cite{Haigh2009}.

\section{ACKNOWLEDGMENTS}

We thank P. DeGiovanni, F. Faure, F. Piazza, A. Smerzi, D. Spenher, T. Roscilde for useful discussions.
We acknowledge financial support from CNRS and from the European MIDAS project.


\begin{thebibliography}{33}
\bibitem{nota}
In typical experiments the system is destroyed after a measurement of $\hat{J_r}$.
Therefore, one should model the process by a Positive Operator Valued Measure (POVM)
rather than by a projective measurement \cite{Nielsen_Chuang}. However, here we are only
interested in the outcomes $r$. These are predicted to be the same for both  POVM and
projective measurement, even if the latter does not describe properly the state of the
system after the measurement.
\end{thebibliography}
\end{document}